\documentclass{article}

\usepackage{amsmath}
\usepackage{amsfonts}
\usepackage{amsthm}
\usepackage{verbatim}

\newcommand{\lb}{\left (}
\newcommand{\rb}{\right )}
\newcommand{\lsb}{\left [}
\newcommand{\rsb}{\right ]}
\newcommand{\lab}{\left \langle}
\newcommand{\rab}{\right \rangle}

\newcommand{\eq}[1]{\begin{equation} #1 \end{equation}}
\newcommand{\eqna}[1]{\begin{eqnarray} #1 \end{eqnarray}}
\newcommand{\be}{\begin{eqnarray}}
\newcommand{\ee}{\end{eqnarray}}
\newcommand{\bc}{}

\newcommand{\p}{\partial}
\newcommand{\mF}{\mathcal{F}}
\newcommand{\e}{\epsilon}

\newcommand{\wZ}{\widetilde{Z}}
\newcommand{\wF}{\widetilde{F}}
\newcommand{\wG}{\widetilde{G}}

\hoffset= - 1.0in         

\title{{\bf Resolvents and Seiberg-Witten representation for Gaussian $\beta$-ensemble} \vspace{.2cm}}
\author{{\bf A.Mironov}\footnote{ {\small {\it
Lebedev Physics Institute} and {\it ITEP, Moscow, Russia}};
mironov@itep.ru; mironov@lpi.ru}, {\bf A.Morozov}\thanks{{\small
{\it ITEP, Moscow, Russia}}; morozov@itep.ru},
{\bf A.Popolitov}\thanks{{\small
{\it ITEP, Moscow, Russia}}; popolit@itep.ru},
{\bf Sh.Shakirov}\thanks{{\small
{\it Department of Mathematics, University of California, Berkeley, USA} and
{\it ITEP, Moscow, Russia}}; shakirov@itep.ru}\date{ }}

\begin{document}
 \maketitle

\vspace{-5.0cm}

\begin{center}
\hfill FIAN/TD-03/11\\
\hfill ITEP/TH-07/11\\
\end{center}

\vspace{3.5cm}

\centerline{ABSTRACT}

\bigskip

{\footnotesize
The exact free energy of matrix model always obeys the
Seiberg-Witten (SW) equations on a complex curve
defined by singularities of the quasiclassical resolvent.
The role of SW differential is played by the exact one-point
resolvent.
We show that these properties are preserved in generalization
of matrix models to beta-ensembles.
However, since the integrability and Harer-Zagier topological
recursion are still unavailable for beta-ensembles,
we need to rely upon the ordinary AMM/EO recursion to evaluate
the first terms of the genus expansion.
Consideration in this paper is restricted to the Gaussian model.
}


\bigskip

\bigskip

\section{Introduction}

Seiberg-Witten (SW) prepotentials ${\cal F}(\vec a)$ \cite{SWfirst,SW2,SW,SWrev}
are defined from the peculiar set of implicit equations:
\be
\left\{\begin{array}{c}
\vec a = \oint_{\vec A} \Omega \\
\\
\frac{\partial {\cal F}(\vec a)}{\partial \vec a}
= \oint_{\vec B} \Omega
\end{array}\right.
\label{SWeq}
\ee
Here $\Omega$ is an $(m,0)$ analytic form
(holomorphic, meromorphic or even possessing essential
singularities) on a family of $d=2m$ complex manifolds with
a system of conjugated $(m,0)$-cycles $\vec A$ and $\vec B$.
When the system (\ref{SWeq}) is resolvable
(its consistency is guaranteed by the Riemann identities)
then $\vec a$ are called flat coordinates on the
moduli space of the family (or simply the flat moduli),
and ${\cal F}(\vec a)$ is a "quasiclassical"
or Whitham $\tau$-function, on this space,
satisfying a set of the (generalized) WDVV equations
(usually as a consequence of the "residue formula") \cite{WDVV}.
This is by now a classical branch of science,
presented in big detail in
numerous papers.

A little more recently it has been realized that
despite the "quasiclassical" nature of the SW equations,
they perfectly survive various "quantization" procedures.
The true conceptual meaning of this phenomenon still
lacks understanding, but the very
fact is getting established more and more reliably.
The latest example is the Bohr-Sommerfeld representation
\cite{MMbz,MMM}
of the LMNS free energy \cite{LMNS}
in the Nekrasov-Shatshvili limit \cite{NSh}
$\epsilon_2=0$:
if also $\epsilon_1=0$, then this free energy is just
the ordinary SW prepotential of \cite{SWfirst,SW},
but remarkably eqs.(\ref{SWeq}) survive when at least
the first "quantization" parameter $\epsilon_1$
is switched on.
Actually it is claimed in \cite{MMSh1,MMShtowaproof}
that they will survive even further:
when both $\epsilon_1$ and $\epsilon_2$ are non-vanishing.
And this claim is inspired by the AGT relations
\cite{AGT,AGTproof,AGTmamoproof},
which provide a matrix model representation of the
LMNS partition function \cite{AGTmamo,MMSh1}.
Then one can use the previous fact of the fundamental
importance: that the exact matrix model free energies
possess the SW representation with the role of SW
differential played by the one-point resolvent
\be
\Omega^{MM}(z) = \rho_1(z) = \left<{\rm Tr}\frac{dz}{z-M}\right>_{MM}
\label{OmeMM}
\ee
which is a meromorphic differential on the
spectral curve $\Sigma^{MM}$.
Again, the SW representation is a kind of straightforward
for the planar free energy
(where it is actually discussed since \cite{DV,DVmore})
but, remarkably, it survives when all higher-genus
corrections in powers of the string coupling constant $g_S$
(i.e. the t'Hooft's coupling $\Lambda = gN$)
are switched on.
This fact is still less known and under-appreciated.
It was mentioned in passing in \cite{AMMfirst}
and in \cite{EOrev},
but its real significance can be illustrated by the
recent suggestion of \cite{MMShtowaproof} to use it in a
conceptual proof of the AGT relations:
by applying the topological recursion procedures \cite{AMMfirst,toprec}
to construct a double deformation of the original
SW free energy, to $g_s = \sqrt{-\epsilon_1\epsilon_2}\neq 0$
and to $\beta = b^2 = -\epsilon_1/\epsilon_2\neq 1$.

It is the goal of the present paper to provide more
{\it illustrations} to the SW representation of exact matrix
model free energies and to make this crucially
important technique more understandable and useable.
The paper is dedicated
entirely to this issue and we avoid mixing it with the
other subjects.
The first illustration of this kind was already provided
in the Appendix to \cite{MMShtowaproof},
we reproduce it here and extend to non-unity $\beta$.
 We do not address non-Gaussian models here, since this requires usage of rather heavy techniques, but we will address this question in forthcoming papers.
Of course, for AGT applications one needs an essentially
non-Gaussian $\beta$-ensemble: the open-contour Dotsenko-Fateev integral
{\it a la} \cite{MMSh1,MMShDF}, which we do not consider in the
present text.
However, the SW representation undoubtedly exists there as well,
also for arbitrary $\beta$ and in all orders of the genus
expansion.

\section {The case of $\beta = 1$: a source of questions and educated guesses}
\label{sec_beta_eq_1}
The partition function is defined as
\eq {
    Z (N)= \frac{1}{N!} \int d \lambda_1 \dots d\lambda_N \prod_{i<j} \left|
    \lambda_i-\lambda_j\right|^{2\beta} e^{-\frac{1}{2g}\sum_i \lambda_i^2}
}
For $\beta=1$ it is equal to
\eq {
    Z (N) = \sqrt{2\pi}^N \sqrt{g}^{N^2} \prod_{k=1}^{N-1} k!
}
Hence, for the free energy $F = \ln Z$ one has (ignoring the
terms quadratic and linear in $N$)
\eq {
    F (N) = \sum_{k=1}^{N-1} \ln (k!)
}

It turns out \cite{GMAMO} that $Z(N)$ is a Toda-chain $\tau$-function and
$F(N)$ possesses the Seiberg-Witten representation (\ref{SWeq}).

That is, let $\rho_1(z)$ be the one-point resolvent of the model
\eq {
    \rho_1 (z) = \lab \sum_i \frac{1}{z-\lambda_i} \rab
}

Then the system of partial differential SW-equations
\eq {
    -\frac{1}{2\pi i} \oint_A \rho_1 (z)dz = a\
    \ \ \ \ \ \ -\oint_B \rho_1(z)dz = \frac{\p \mF_{SW}}{\p a},
}
determines the SW prepotential, which, as one can check using the explicit expression for
the resolvent from \cite{MMShtowaproof}, is equal to the free energy
\eq {
    \mF_{SW} (N) = F (N)
    \label{SW_representation}
}
and this equality just gives the SW-representation of free energy
of the matrix model.

\vspace{5 mm}

Another remarkable fact is that the one-point resolvent satisfies the difference equation
\cite{HZ,ShMMHarerZagier}
\eq {
    \rho_1 (N+1,z) + \rho_1 (N-1,z) - 2\rho_1 (N,z) = \frac{\p^2}{\p z^2} \rho_1 (N,z),
    \label{res_eqn}
}
which implies that its B-periods satisfy \cite{MMShtowaproof}
\eq {
    \Pi_B (N+1) + \Pi_B (N-1)-2 \Pi_B (N) = -\frac{1}{N}
    \label{per_eqn}
}

Eq.(\ref{res_eqn}) is closely related to integrability of $Z(N)$, that is,
to the Toda chain equation \cite{GMAMO}
\eq {
    Z(N) \p_1^2 Z(N) - \p_1 Z(N) \p_1 Z(N) = Z(N+1) Z(N-1),
}
where $\p_1 Z(N) = \lab \sum_i \lambda_i \rab$ and $\p_1^2 Z(N) = \lab \lb \sum_i \lambda_i\rb^2 \rab$.

Eq.(\ref{per_eqn}) is a weaker corollary of (\ref{res_eqn}).

Knowing these facts, the following questions arise naturally:
\begin{itemize}
 \item Does (\ref{SW_representation}) hold as well in the $\beta \neq 1$ case?
 \item Is there some $\beta$-deformed version of (\ref{res_eqn}) and (\ref{per_eqn})?
\end{itemize}

The rest of the paper is devoted to the affirmative answer to the first question.
A partial progress in answering the second one is outlined in the Appendix.

\section{Resolvents}
\label{sec_resolvents}
\subsection{Ward identities: generalities}

A powerful technique for evaluating correlators in matrix models is known under
different names: of the Virasoro constraints, of the loop equations, of the Ward
identities \cite{Vir,AMMfirst}. It relies on "the general covariance" of partition function: that is,
the invariance of integral under
arbitrary change of integration variables. For the
eigenvalue model, not obligatory Gaussian, the Virasoro constraints can be deduced
as follows \cite{confmm}. Consider the obvious identity
\eq {
   \sum_k \int d\lambda_1\dots d\lambda_N \frac{\p}{\p \lambda_k} \lb \lambda_k^n \Delta^{2\beta} e^{-\frac{1}{g} \sum_i V(\lambda_i)} S_{i_1}\dots S_{i_m} \rb = 0,
}
where $S_i = \sum_a \lambda_a^i$ and $\Delta$ is the absolute value of the
Van-der-Monde determinant.
Here $V(\lambda) = \sum_k T_k \lambda^k$; in the Gaussian case only $T_2 = 1/2$ is
non-vanishing.

One can easily check that
\eq {
  \sum_k  \frac{\p}{\p \lambda_k} \lb \lambda_k^n \Delta^{2\beta} \rb = \lb \beta \sum_{a=0}^{n-1} S_a S_{n-1-a} + (1-\beta) n S_{n-1} \rb \Delta^{2\beta}
}
and this is the only piece of equation that changes when one changes $\beta$.

Differentiation of the potential term gives
\eq {
  \sum_k  \lambda_k^n \frac{\p}{\p \lambda_k} \lb e^{-\frac{1}{g} \sum_i V(\lambda_i)}\rb = \lb -\frac{1}{g} \sum_a V'(\lambda_a) \lambda_a^n \rb e^{-\frac{1}{g} \sum_i V(\lambda_i)}
}
and this is the only model-dependent part of our consideration.

Differentiation of the remaining terms gives
\eq {
  \sum_k  \lambda_k^n \frac{\p}{\p \lambda_k} \lb S_{i_1} \dots S_{i_m} \rb = \sum_{j=1}^m i_j S_{i_1}\dots S_{i_j+n-1} \dots S_{i_m}
}

\vspace{5 mm}

Now, having all the ingredients of the equations, one can write them in various forms.

\paragraph {Virasoro constraints.}
If one denotes the disconnected correlator as
\eq {
    C_{i_0,., i_m}  = \lab S_{i_0} \dots S_{i_m}\rab
}
the above considerations imply that
\eq {
    \beta \sum_{a=0}^{n-1} C_{a,n-1-a,i_1,., i_m} + (1-\beta) n C_{n-1,i_1,., i_m} - \frac{1}{g} \sum_k kT_k C_{n-1+k,i_1,., i_m} + \sum_{j=1}^m i_j C_{i_1,., i_j+n-1,., i_m} = 0
    \label{virasoro_constraints}
}

\paragraph{Differential ($\widetilde{W}$) operators.}
If one works with the generic partition function (with infinitely many non-fixed
times) one can write these equations as a differential equation on the (full)
partition function. Namely, let the potential have the form
\eq {
    V(\lambda) = (T_0 + t_0)N + \sum_{k=1}^\infty (T_k + t_k) \lambda^n,
}
where $T_k$ are background  values of source fields (usually, only finitely many
of them are non-zero) and $t_k$ are perturbations of these background values.
The partition function is thought of as a formal series in $t_k$.
Note that, for the non-normalized average, one has
\eq {
    \lab S_a \rab = -g \frac{\p}{\p t_a} \lab 1 \rab
}
and, hence, the Virasoro constraints (\ref{virasoro_constraints}) can be written as
\eq {
      \frac{\p}{\p t_{i_1}} \dots \frac{\p}{\p t_{i_m}} \lb
\sum_{k=0}^\infty k (T_k+t_k) \frac{\p}{\p t_{k-1+n}}
     + g\lb 1-\beta \rb n \frac{\p}{\p t_{n-1}} +
     g^2 \beta \sum_{a=1}^{n-1} \frac{\p^2}{\p t_a \p t_{n-1-a}} \rb
        Z = 0
}

\paragraph{Loop equations.}
They equations arise when one sums up all the Virasoro constraints with
the weights
$\frac{1}{z^{n+1}}$ and writes the resulting equation in terms of the resolvents. For
this purpose, it is convenient to rewrite the Van-der-Monde part of the identity as

\eq {
   \sum_k \frac{\p}{\p \lambda_k} \lb \lambda_k^n \Delta^{2\beta} \rb = \lb 2\beta \sum_{i<j} \frac{\lambda_i^n-\lambda_j^n}{\lambda_i-\lambda_j} + \sum_a n \lambda_a^{n-1}\rb \Delta^{2\beta}
}

Now summing up all the contributions one gets
\be
    \label{loop_eq}
     \beta r\lb z_0,z_0,z_1,.,z_m\rb + (\beta-1) \frac{\p}{\p z_0}
r(z_0,z_1,.,z_m) + \sum_{j=1}^m \frac{\p}{\p z_j}
\frac{r(z_1,.,z_m) - r(z_1,.,z_0,.,z_m)}{z_m-z_0} -\\
-\frac{1}{g} \sum_{k=0}^\infty k T_k z_0^{k-1} r(z_0,z_1,.,z_m)
+ \frac{1}{g} \sum_{k=0}^\infty k T_k \sum_{j=0}^{k-2} z_0^j
\frac{1}{2\pi i}\oint_\infty dz z^{k-2-j} r(z,z_1,.,z_m)
     = 0 \nonumber,
\ee
where $r(z_0,.,z_m)$ is the disconnected resolvent
\eq {
    r(z_0,.,z_m) = \lab \sum_{i_0} \frac{1}{z_0 - \lambda_{i_0}} \dots
\sum_{i_m} \frac{1}{z_m - \lambda_{i_m}}\rab
}

To solve these equations perturbatively in $g$ one has to rewrite the disconnected
resolvents in terms of the connected ones. Then, the iteration procedure becomes
well-defined: at each step of the procedure one has a system of linear equations for
$\rho_{i,j}$, with fixed value of $i+j$. The expansion in powers $k$ of $g$, as usual,
counts contributions of genus $k/2$ Riemann surfaces in string (or topological)
expansion. Here $\rho_{i,j}$ stands for the genus $j$ contribution to the $i$-point
connected resolvent.

\subsection{Prerequisite: particular correlators}

The Ward identities in the form of the Virasoro constraints are very
helpful in evaluating individual correlators $C_{i_1\dots i_m}$. The
advantage of this method is that the answers are exact in $g$ and
one may not rewrite the disconnected correlators in terms of the
connected ones for the iteration procedure to work (this simplifies
the work drastically if one uses symbolic computer computations).

\vspace{5 mm}

To give an impression of what individual correlators look like we provide the first
few one- and two-point correlators. Note that $K$
denotes the \textit{connected} correlators, and $\Lambda \equiv N g$.

\be
\label{expl_correls_1}
K_k = &C_k=\lab \sum_i \lambda_i^k \rab = \lab\lab \sum_i \lambda_i^k
\rab\rab & \\
K_0 = &\Lambda  \nonumber\\
K_2 = &\Lambda  (\beta  \Lambda -\beta +1)  \nonumber\\
K_4 = &\Lambda  \left(2 \beta ^2 \Lambda ^2-5 \beta ^2 \Lambda +3 \beta ^2+5
\beta  \Lambda -5 \beta +3\right) & \nonumber\\
K_6 = &5 \beta ^3 \Lambda ^4+\left(22 \beta ^2-22 \beta ^3\right) \Lambda
^3+\left(32 \beta ^3-54 \beta ^2+32 \beta \right) \Lambda ^2+\left(-15 \beta
^3+32 \beta ^2-32 \beta +15\right) \Lambda & \nonumber\\
K_8 = &14 \beta ^4 \Lambda ^5+\left(93 \beta ^3-93 \beta ^4\right) \Lambda
^4+\left(234 \beta ^4-398 \beta ^3+234 \beta ^2\right) \Lambda ^3
& \nonumber \\ &+\left(-260
\beta ^4+565 \beta ^3-565 \beta ^2+260 \beta \right) \Lambda ^2+\left(105 \beta
^4-260 \beta ^3+331 \beta ^2-260 \beta +105\right) \Lambda &\nonumber \\
& \dots \nonumber &
\ee

\be
\label{expl_correls_2}
K_{k,j} = & C_{k,j}-C_kC_j=\lab\lab \sum_i \sum_l \lambda_i^k \lambda_l^j \rab\rab \\
K_{1,1} = & \Lambda \nonumber\\
&&\nonumber \\
K_{1,3} = & 3 \Lambda  (\beta  (\Lambda -1)+1) \nonumber\\
K_{2,2} = & 2 \Lambda  (\beta  (\Lambda -1)+1) \nonumber \\
&&\nonumber \\
K_{1,5} = & 10 \beta ^2 \Lambda ^3+5 \left(5 \beta -5 \beta ^2\right)
\Lambda ^2+5 \left(3 \beta ^2-5 \beta +3\right) \Lambda \nonumber \\
K_{2,4} = & 4 \Lambda  (\beta  (\Lambda -1) (\beta  (2 \Lambda -3)+5)+3)
\nonumber \\
K_{3,3} = & 3 \Lambda  (\beta  (\Lambda -1) (\beta  (4 \Lambda -5)+9)+5)
\nonumber \\
& \dots \nonumber &
\ee

In terms of \textbf{the CFT-inspired variables} $M = b \Lambda$ and $Q =
b-\frac{1}{b}$, $b = \sqrt{\beta}$ these read

\be
K_0 =&\frac{M}{b} \\
K_2 =&M  (M-Q) \nonumber\\
K_4 =&Mb (1+2 M^2-5 M Q+3 Q^2) \nonumber\\
K_6 =&M b^2 (5 M (2+M^2)-(13+22 M^2) Q+32 M Q^2-15 Q^3) \nonumber\\
K_8 =&M b^3 (21+14 M^4-93M^3 Q+160 Q^2+105 Q^4-5 M Q (43+52 Q^2)+M^2 (70+234
Q^2)) \nonumber \\
& \dots \nonumber &
\ee

\be
K_{1,1} = & \frac{M}{b} \\
\nonumber \\
K_{1,3} = & 3 M (M-Q) \nonumber \\
K_{2,2} = & 2 M (M-Q) \nonumber \\
\nonumber \\
K_{1,5} = & 5 M b (1+2 M^2-5 M Q+3 Q^2) \nonumber \\
K_{2,4} = & 4 M b (1+2 M^2-5 M Q+3 Q^2)\nonumber \\
K_{3,3} = & 3 M b (1+4 M^2-9 M Q+5 Q^2) \nonumber \\
& \dots \nonumber &
\ee

Note the remarkable simplification in comparison with (\ref{expl_correls_1}) and
(\ref{expl_correls_2}).

\subsection {The answer for resolvent at $\beta = 1$}
Just for completeness (and in part to emphasize
the relative complexity of the $\beta \neq 1$ case) we begin from the well-known
one-point resolvent at $\beta = 1$ \cite{AMMfirst}:
\eq {
    \rho_1 = \lab \sum_i \frac{1}{z - \lambda_i}\rab = \sum_{k = 0}^\infty \rho_{1,k} g^k
}

The particular genus contributions are
\eq {
    \rho_{1,0} \lb z \rb = \frac{1}{2} \lb z - y(z) \rb \nonumber
}

\eq {
    \rho_{1,2} \lb z \rb = \frac{\Lambda}{y(z)^5}
}

\eq {
    \rho_{1,4} \lb z \rb = \frac{21\Lambda (\Lambda + z^2)}{y(z)^{11}}
\nonumber
}
\eq {
    \rho_{1,6} \lb z \rb = \frac{11 \Lambda (158 \Lambda^2 + 558 \Lambda
z^2 + 135 z^4)}{y(z)^{17}} \nonumber,}
\eq{\ldots
}
where $y(z)^2 = z^2 - 4 \Lambda$ and all $\rho_{1,2k+1}$ vanish. General formulae for
$\rho_{1,2n}$ can be obtained by integral transformation from exact Harer-Zagier functions,
see \cite{HZ,AMMfirst,ShMMHarerZagier}.

\subsection {The answer for $\rho_1$ at generic $\beta \neq 1$}
\label{sec_res_beta}

The loop equations (\ref{loop_eq}) in the case of Gaussian model acquire
a very simple form
\be
\beta r\lb z_0,z_0,z_1,.,z_m\rb + (\beta-1) \frac{\p}{\p z_0}
r\lb z_0,z_1,.,z_m\rb +
\sum_j \frac{\p}{\p z_j}
\frac{r\lb z_1,.,z_j,.,z_m \rb - r \lb z_1,.,z_0,.,z_m
\rb}{z_j-z_0} - \\
-\frac{1}{g} z_0 r \lb z_0,z_1,.,z_m\rb + \frac{\Lambda}{g^2} r\lb
z_1,.,z_m\rb = 0 \nonumber
\ee
where $r$ denotes the disconnected resolvent
\eq {
    r \lb z_1,.,z_m\rb = \lab
\sum_{i_1}\frac{1}{z_1-\lambda_{i_1}}\dots\sum_{i_m}\frac{1}{z_m-\lambda_{i_m
} } \rab
}

To solve this system of equations one should rewrite the disconnected correlators
in terms of the connected ones and substitute the connected correlators by their
Laurent expansion \cite{ShMMCIV-DV}.

Thus, assuming that $\rho_1\lb z\rb = \frac{1}{g} \sum_{i=0}^\infty
\rho_{1,i}\lb z\rb \cdot g^i$ (so that the even parts of $\rho$ are associated with
oriented surfaces, while the odd parts with the non-oriented ones, with half-integer
genera), one gets for the first few terms:
\eq {
    \rho_{1,0} \lb z \rb = \frac{z}{2 \beta }-\frac{y(z)}{2 \beta } = \frac{1}{2\beta} \lb z - y(z) \rb
\nonumber
}

\eq {
    \rho_{1,1} \lb z \rb = \frac{\frac{1}{2}-\frac{1}{2 \beta
}}{y(z)}+\frac{\frac{z}{2 \beta }-\frac{z}{2}}{y(z)^2}  = \frac{\beta-1}{2\beta y(z)} \lb 1 - \frac{z}{y(z)}\rb \nonumber
}

\eq {\label{32}
    \rho_{1,2} \lb z \rb =\frac{5 \beta ^2 \Lambda -9 \beta  \Lambda +5
\Lambda }{y(z)^5}+\frac{\beta +\frac{1}{\beta }-2}{y(z)^3}+\frac{-\beta
z-\frac{z}{\beta }+2 z}{y(z)^4}
}

\eqna {
    \rho_{1,3} \lb z \rb = &
(\beta-1) \lb \frac{10-19\beta + 10\beta^2}{2\beta y(z)^5} \lb 1 - \frac{z}{y(z)}\rb + \frac{5\Lambda(5 -9\beta +5\beta^2)}{y(z)^7}
+ \frac{-\Lambda z (30 - 43\beta + 30\beta^2)}{y(z)^8} \rb \nonumber
}

\eqna {
    \rho_{1,4} \lb z \rb = & \frac{37 \beta ^3-\frac{273 \beta ^2}{2}+199
\beta +\frac{37}{\beta }-\frac{273}{2}}{y(z)^7}+\frac{-37 \beta ^3 z+\frac{273
\beta ^2 z}{2}-199 \beta  z-\frac{37 z}{\beta }+\frac{273
z}{2}}{y(z)^8}+\frac{419 \beta ^4 \Lambda -1357 \beta ^3 \Lambda +1897 \beta ^2
\Lambda -1357 \beta  \Lambda +419 \Lambda }{y(z)^9} &\nonumber\\& +\frac{-240
\beta ^4 \Lambda z+824 \beta ^3 \Lambda  z-1168 \beta ^2 \Lambda  z+824 \beta
\Lambda z-240 \Lambda  z}{y(z)^{10}}+\frac{1105 \beta ^5 \Lambda ^2-3240 \beta
^4 \Lambda ^2+4375 \beta ^3 \Lambda ^2-3240 \beta ^2 \Lambda ^2+1105 \beta
\Lambda ^2}{y(z)^{11}}
\nonumber
}

\eqna {
    \rho_{1,5} \lb z \rb = &
(\beta - 1) \Big [
\frac{(706 - 2379 \beta +3367\beta^2 - 2379 \beta^3 + 706 \beta^4)}{2\beta y^9(z) \lb 1 - \frac{z}{y(z)}\rb} +
\frac{4351 - 13458\beta + 18508\beta^2 - 13458\beta^3 + 4351\beta^4}{y(z)^{11}} \nonumber & \\ &
- \frac{3\Lambda z(1530 - 4241 \beta + 5764 \beta^2 - 4241 \beta^3 + 1530\beta^4)}{y(z)^{12}}
+ \frac{55\beta\Lambda^2(221 - 648\beta + 875\beta^2 -648\beta^3 +221\beta^4)}{y(z)^{13}} \nonumber & \\ &
- \frac{4\beta\Lambda^2 z (3390 -7883\beta + 10420 \beta^2 - 7883\beta^3 +3390\beta^4)}{y(z)^{14}}
\Big ] \nonumber
}

\eqna {
    \rho_{1,6} \lb z \rb = &\frac{4081 \beta ^5-\frac{40405 \beta
^4}{2}+44699 \beta ^3-57155 \beta ^2+44699 \beta +\frac{4081}{\beta
}-\frac{40405}{2}}{y(z)^{11}}+\frac{-4081 \beta ^5 z+\frac{40405 \beta ^4
z}{2}-44699 \beta ^3 z+57155 \beta ^2 z-44699 \beta  z-\frac{4081 z}{\beta
}+\frac{40405 z}{2}}{y(z)^{12}}&\nonumber\\ &+\frac{77597 \beta ^6 \Lambda
-340402 \beta ^5 \Lambda +702694 \beta ^4 \Lambda -878293 \beta ^3 \Lambda
+702694 \beta ^2 \Lambda -340402 \beta  \Lambda +77597 \Lambda
}{y(z)^{13}}&\nonumber\\ &+\frac{-59040 \beta ^6 \Lambda  z+269328 \beta ^5
\Lambda  z-564000 \beta ^4 \Lambda  z+707424 \beta ^3 \Lambda  z-564000 \beta
^2 \Lambda z+269328 \beta  \Lambda  z-59040 \Lambda  z}{y(z)^{14}}&\nonumber\\
&+\frac{451720 \beta ^7 \Lambda ^2-1792898 \beta ^6 \Lambda ^2+3483419 \beta ^5
\Lambda ^2-4266464 \beta ^4 \Lambda ^2+3483419 \beta ^3 \Lambda ^2-1792898
\beta ^2 \Lambda ^2+451720 \beta  \Lambda ^2}{y(z)^{15}}&\nonumber\\
&+\frac{-189840 \beta ^7 \Lambda ^2 z+821128 \beta ^6 \Lambda ^2 z-1656256
\beta ^5 \Lambda ^2 z+2049936 \beta ^4 \Lambda ^2 z-1656256 \beta ^3 \Lambda ^2
z+821128 \beta ^2 \Lambda ^2 z-189840 \beta  \Lambda ^2
z}{y(z)^{16}}&\nonumber\\ &+\frac{828250 \beta ^8 \Lambda ^3-3012930 \beta ^7
\Lambda ^3+5531740 \beta ^6 \Lambda ^3-6644070 \beta ^5 \Lambda ^3+5531740
\beta ^4 \Lambda ^3-3012930 \beta ^3 \Lambda ^3+828250 \beta ^2 \Lambda
^3}{y(z)^{17}} \nonumber,
}
here $y\lb z\rb^2 = z^2 - 4\Lambda\beta$ defines the spectral curve,
which in this case is the torus with a degenerated handle (located at infinity of the
complex plane).

\paragraph{$\beta \rightarrow \frac{1}{\beta}$ symmetry.}
The AGT relation implies that the $\beta$-deformed matrix model should be related
to some CFT, with the central charge of the corresponding CFT given by
\eq {
    c = 1 - 6 \lb \sqrt{\beta} - \frac{1}{\sqrt{\beta}}\rb^2
}
This hints that there should be the symmetry $\beta \rightarrow \frac{1}{\beta}$
present
in the matrix model despite this is far from obvious in the original expression
(3). And, indeed, one can see that if one rescales the quantities
in the following way:
\eq {
    z' = \sqrt{\beta} z,\ \ \ \ \  \rho'_{1,g} = \sqrt{\beta}^{g+1} \rho_{1,g},
}
the resulting $\rho'_{1,g}$ are symmetric w.r.t. $\beta \rightarrow \frac{1}{\beta}$.

\section {Seiberg-Witten construction}
\label{sec_SW}
\subsection{Ideology}
The Seiberg-Witten construction, originally proposed to obtain the low-energy
effective action in $\mathcal{N} = 2$ SUSY gauge theory is in fact a manifestation of
a more general statement.

The starting objects in the SW representation are the algebraic curve and the meromorphic
differential $\lambda_{SW}$ on it. Given such a data, one writes the following system
of equations
\eq {
    \oint_{A_i} \lambda_{SW} \sim  a_i\ \ \
    \ \ \ \ \ \oint_{B_i} \lambda_{SW} \sim \frac{\p \mF_{SW}}{\p a_i},
}
where $A_i$ and $B_i$ form a symplectic basis of cycles on the algebraic curve and
proportionality coefficients in equations slightly depends on setting.

It turns out that a huge source of the SW data is provided by the eigenvalue models
(EVM).
Namely, the algebraic curve is the spectral curve of the given EVM, while the SW
differential is $\rho_1(z)dz$, $\rho_1$ being the one-point resolvent. Note that
the original SW construction corresponds to the zeroth order of genus expansion of
the resolvent in $g$, and taking into account further terms of the expansion
corresponds to the deformation (quantization) of the original SW differential and
prepotential. Remarkably, the \textit{all genus} free energy, not only its genus zero
part, continues to satisfy the SW equations.

We fix the proportionality coefficients in the SW equations as follows
\eq {
    - \frac{1}{2\pi i} \oint_{A_i} \rho_1(z)dz  =  a_i\ \
    \ \ \ \ \ \ -\beta \oint_{B_i} \rho_1(z)dz = \frac{\p \mF_{SW}}{\p a_i},
    \label{SWEqnsPrecise}
}
as relation between the free energy and the SW prepotential is most transparent
in this way. Note that $\beta$ appeared as a coefficient in the second equation, \cite{MMM}.

\subsection {Calculation of A- and B-periods}
 Now let us apply the SW construction to $\rho_1$ that we found in section
 \ref{sec_res_beta}.

The spectral curve is given by the equation
 \eq {
     y^2 = z^2 - 4\Lambda\beta
}
 The A-cycle encircles the ramification points $-\sqrt{4\Lambda\beta}$ and
 $\sqrt{4\Lambda\beta}$, while the B-cycle encircles $\sqrt{4\Lambda\beta}$ and
 $\infty$.

 Since in this case the value of A-period is equal to the residue at infinity, the
 A-period gets contributions only from $\rho_{1,0}$ and $\rho_{1,1}$:
 \eq {
    a = -\frac{1}{2\pi i} \oint_{-\sqrt{4\Lambda\beta}}^{\sqrt{4\Lambda\beta}}
    \rho(z) dz = N + \frac{1-\beta}{2\beta}
}
Note at this point that the dependence $a(N)$ is \textit{linear} and one can safely
substitute $\frac{\p}{\p a}$ by $\frac{\p}{\p N}$ in the SW equation to simplify
calculations.

Evaluating the B-periods is more tricky, the following formula is of great use
\eq {
    \oint_{\sqrt{4\Lambda\beta}}^{+\infty} \frac{dz}{y(z)^p} = \frac{1}{2^{2p-3}
\lb\Lambda\beta\rb^{(p-1)/2}} \frac{\Gamma(p-1)\Gamma(1-p/2)}{\Gamma(p/2)}
}
To deduce this formula, one has to make the change of variables $z =
\frac{2-\zeta}{\zeta}\sqrt{4\Lambda\beta} $ and notice that the resulting integral
is proportional to the integral representation for the Euler $B$-function
\be
\oint_{\sqrt{4 \beta \Lambda}}^{+\infty} \lb z^2 - 4 \beta \Lambda \rb^{-p/2} dz =
\lb 4 \beta \Lambda \rb^{-p/2+1/2} \oint_1^{+\infty} \lb w^2 - 1 \rb^{-p/2} dw = \\
=\lb 4 \beta \Lambda \rb^{-p/2+1/2} 4^{-p/2} 2 \oint_0^1 (1-\zeta)^{-p/2} \zeta^{p-2} d\zeta =
2^{-2p+3} \lb \beta \Lambda \rb^{-p/2+1/2} \frac{\Gamma\lb1-\frac{p}{2}\rb \Gamma\lb p-1
\rb}{\Gamma \lb \frac{p}{2}\rb} \nonumber
\ee
The terms in (\ref{32}) with odd powers of $z$ do not contribute to the periods, since they are
total derivatives. For instance,
\eq {
    \oint_{\sqrt{4\Lambda\beta}}^{+\infty} \frac{z dz}{y(z)^p} = -\frac{1}{p-2} \oint_{\sqrt{4\Lambda\beta}}^{+\infty} d \lb \frac{1}{y(z)^{p-2}}\rb = 0
}
Note that this is a contour integral and the contour does not pass through the singularities
of the integrand.
\vspace{5mm}

Alternatively, one may exploit the fact that $y(z)$ satisfies the differential equation
\eq {
    \frac{\p}{\p \Lambda} y^p= -2\beta p y^{p-2},
}
and so do its B-periods. Together with the initial conditions
\eq {
    \oint_B \frac{dz}{y(z)} = - \ln \beta \Lambda
}
and
\eq {
    \oint_B y^p dz = 0\Big{|}_{\Lambda = 0}; \ \ \ \ p \neq -1,
}
this gives (only minor modifications occur in comparison with $\beta = 1$ case)

\begin{center}
\begin{tabular}{|c|c|c|}
\hline
 $n$ & $\oint_B y^n$ & $\oint_B y^{-n}$ \\ \hline
 1 & $-2 \beta  (\Lambda -\Lambda  \log (\beta  \Lambda ))
$ & $-\log (\beta  \Lambda )$
\\ \hline
3 & $-6 \beta  \left(\beta  \Lambda ^2 \log (\beta  \Lambda )-\frac{3 \beta
\Lambda
^2}{2}\right)$
& $-\frac{1}{2 \beta  \Lambda }$
\\ \hline
5 & $-10 \beta  \left(\frac{11 \beta ^2 \Lambda ^3}{3}-2 \beta ^2 \Lambda ^3
\log
(\beta  \Lambda )\right)$
& $\frac{1}{12 \beta ^2 \Lambda ^2}$
\\ \hline
7 & $-14 \beta  \left(5 \beta ^3 \Lambda ^4 \log (\beta  \Lambda )-\frac{125
\beta
^3 \Lambda ^4}{12}\right)$
& $-\frac{1}{60 \beta ^3 \Lambda ^3}$
\\ \hline
9 & $-18 \beta  \left(\frac{959 \beta ^4 \Lambda ^5}{30}-14 \beta ^4 \Lambda ^5
\log
(\beta  \Lambda )\right)$
& $\frac{1}{280 \beta ^4 \Lambda ^4}$
\\ \hline
11 & $-22 \beta  \left(42 \beta ^5 \Lambda ^6 \log (\beta  \Lambda )-\frac{1029
\beta
^5 \Lambda ^6}{10}\right)$
& $-\frac{1}{1260 \beta ^5 \Lambda ^5}$
\\ \hline
13 & $-26 \beta  \left(\frac{11979 \beta ^6 \Lambda ^7}{35}-132 \beta ^6
\Lambda ^7
\log (\beta  \Lambda )\right)$
& $\frac{1}{5544 \beta ^6 \Lambda ^6}$
\\ \hline
\end{tabular}
\end{center}

\vspace{5mm}
Thus, for the B-periods of $\rho_{1,i}$ one gets
(in the case of $\rho_{1,0}$ and $\rho_{1,1}$ one has to evaluate
the integrals for $p = -1+\e$ and $1+\e$ respectively, and then to neglect the
terms which diverge as $\e\rightarrow 0$; since these terms are constant and linear
in $\Lambda$, this is safe)
\eq {
    \oint_B \rho_{1,0}(z)dz = -\Lambda \ln\Lambda \nonumber
}
\eq {
    \oint_B \rho_{1,1}(z)dz = \frac{1-\beta}{2\beta} \ln\Lambda \nonumber
}
\eq {
    \oint_B \rho_{1,2}(z)dz = \frac{-1+3\beta-\beta^2}{12\beta^2\Lambda}
    \label{BPeriodValues}
}
\eq {
    \oint_B \rho_{1,3}(z)dz = \frac{1-\beta}{24\beta^2\Lambda^2} \nonumber
}
\eq {
    \oint_B \rho_{1,4}(z)dz = \frac{1-5\beta^2+\beta^4}{360
\beta^4\Lambda^3} \nonumber
}
\eq {
    \oint_B \rho_{1,5}(z)dz = \frac{-1+\beta^3}{240\beta^4\Lambda^4}
\nonumber
}
\eq {
    \oint_B \rho_{1,6}(z)dz =
\frac{-2+7\beta^2+7\beta^4-2\beta^6}{2520\beta^6\Lambda^5} \nonumber
}
Remarkably, despite the complexity of $\rho_{1,i}$ grows very
fast (exponentially) with increasing $i$, their B-periods complexity increases
not so fast (linearly).

Already at this stage one can see that these formulas agree with the generic ones from
\cite{ShMMCIV-DV}
\eqna {
    \oint_B \rho_{1,2m+2} = \sum_{s=0}^{m+1} B_{2m-2s} B_{2s} \frac{\Gamma(2m+1)}{\Gamma(2s+1)\Gamma(2m-2s+3)} \beta^{-2s} \frac{1}{N^{2m+1}} ,\ m \geq 0 \nonumber\\
    \oint_B \rho_{1,2m+1} = \lb \frac{1}{2\beta} - \frac{1}{2\beta^{2m}}\rb \frac{B_{2m+2} (2m-1)}{(2m+1)(2m+2)}\frac{1}{N^{2m}}, m \geq 1,
}
In \cite{ShMMCIV-DV} they were deduced from eq.(\ref{SWGauss_true}), which we are now aiming to derive.

\subsection {Relation to free energy}

Partition function for the $\beta$-deformed Gaussian eigenvalue model
is defined as

\eq {
    Z (N)= \frac{1}{N!} \int d \lambda_1 \dots d\lambda_N \prod_{i<j} \left|
    \lambda_i-\lambda_j\right|^{2\beta} e^{-\frac{1}{2g}\sum_i \lambda_i^2}
}
and can be calculated explicitly. Generalization of (4) for $\beta\ne 1$ is
(see \cite{ShMMCIV-DV})

\eq {\label{48}
Z (N)= \sqrt{2\pi}^N \sqrt{g}^{\beta N^2 + (1-\beta)N} \prod_{k=1}^{N} \frac{\Gamma(1+\beta k)}{\Gamma(1+\beta)} \cdot \frac{1}{\Gamma(N+1)}
}
Now we are ready to check that the free energy

\eq {
    F (N)= \ln Z \sim \sum_{k=1}^N \ln \Gamma \lb 1+\beta k\rb - \ln N!,
}
is equal to the SW prepotential.

Indeed, one can calculate the $N$-derivative of $F(N)$ and apply the Euler-Maclaurin formula,
see eq.(\ref{EulerMcLaurinFormula}) in the Appendix, to obtain
\eqna {
\frac{\p}{\p N} F \lb \frac{\Lambda}{g} \rb = & \frac{1}{g}\beta \Lambda \ln \Lambda + \frac{\beta-1}{2}\ln\Lambda +
g\frac{1-3\beta+\beta^2}{12 \beta\Lambda} + g^2\frac{\beta - 1}{24\beta \Lambda^2} +
g^3\frac{-1+5\beta^2-\beta^4}{360 \beta^3 \Lambda^3} + &  \nonumber\\
&g^4\frac{1-\beta^3}{240\beta^3 \Lambda^4} +
g^5\frac{2-7\beta^2-7\beta^4+2\beta^6}{2520 \beta^5 \Lambda^5} +
o\lb\frac{1}{\Lambda^5}\rb &  \nonumber
}
this expression can be now compared with (45), taking into account the factor $-\beta$ in
(\ref{SWEqnsPrecise}).
\vspace{5mm}

Finally, one obtains
\eq {
  \boxed{F = \mF_{SW}}
  \label{SWGauss_true}
}
\vspace{-1mm}

This is the main statement of the paper:

\vspace{4mm}

\parbox{15cm}{
\textbf{\large The exact free energy of the
Gaussian $\beta$-ensemble satisfies the SW equations
(\ref{SWEqnsPrecise}) with the exact resolvents in the role of the
SW differential}.
}

\vspace{1cm}

\section*{Acknowledgements}

Our work is partly supported by Ministry of Education and Science of
the Russian Federation under contract 02.740.11.5194, by RFBR
grants 10-02-00509-a (A.Mir. \& A.P.) and 10-02-00499 (A.Mor.\& Sh.Sh.),
by joint grants 11-02-90453-Ukr, 09-02-93105-CNRSL, 09-02-91005-ANF,
10-02-92109-Yaf-a, 11-01-92612-Royal Society.

\appendix
\section{Appendix. Towards understanding of $\beta \neq 1$}

In this Appendix we outline a few topics which are poorly understood but are
of crucial importance for the future theory of $\beta$-ensembles.

\subsection{Integrability}
At $\beta = 1$ we saw that the free energy and the resolvent satisfy integrable
differential-difference equations (\ref{res_eqn}) and (\ref{per_eqn}).
These equations are intimately related with the Toda integrable structure of the
Gaussian matrix model (in general case, it becomes KP integrability).
In particular, the Toda equation can be written as follows
\eq {
    \frac{\p^2}{\p t_1^2} \ln Z(N) =  {Z (N+1) Z(N-1)\over Z^2(N)}
}
and, in terms of the free energy,
\eq {
    F(N+1) - 2F(N)+F(N-1)  = \ln\lb \frac{\p^2}{\p t_1^2} F(N)\rb
}
Then, by differentiating w.r.t. $t_i$ and applying the Virasoro constraints,
one obtains
\eq {
    K_i (N+1) - 2K_i (N) + K_i(N-1) = \frac{i(i-1)}{N} K_{i-2} (N)
}
Summing these equations with the weights $\frac{1}{z^{i+1}}$, one obtains
eq.(\ref{res_eqn}). Eq.(\ref{per_eqn}) then follows from (\ref{res_eqn}) if one
integrates it along the B-period on the spectral curve.
\vspace{5mm}

What is the $\beta$-deformation of Toda/KP integrability is a very important and
intriguing question, but hard to tackle straightforwardly. As we shall see, eq.(\ref{per_eqn})
can be $\beta$-deformed but integrability requires more: a $\beta$-deformation of
(\ref{res_eqn}) which is still unknown.

\subsubsection{Difference equation for periods}
For $\beta=1$ the equation (\ref{per_eqn}) reads
\eq {
    \Pi_B(\Lambda +1) - 2\Pi_B(\Lambda) + \Pi_B(\Lambda -1) = -\frac{1}{\Lambda},
}
where $\Pi_B (\Lambda)$ stands for the B period of $\rho$.

Experimentally one can see (e.g. expanding the l.h.s. into the $\frac{1}{\Lambda}$
series) that, for $\beta \neq 1$, this equation deforms to
\eq {
    \Pi_B \lb \Lambda + \frac{1}{\beta} \rb - \Pi_B \lb \Lambda \rb - \Pi_B \lb \Lambda + \frac{1-\beta}{\beta} \rb + \Pi_B \lb \Lambda - 1\rb = -\frac{1}{\beta \Lambda}
    \label {SW_diff_eqn}
}

\subsubsection{Difference equation for resolvents}
We, however, were unable to find a $\beta$-deformed version of (\ref{res_eqn}), even
the corrections of the first order in $\beta-1$ are missing. What one can say for
sure is that in the required generalization the both sides of (\ref{res_eqn}) deform
stronger than they do in (\ref{SW_diff_eqn}).

\subsection {Harer-Zagier topological recursion}
A detailed description of the Harer-Zagier functions for $\beta = 1$ can be found e.g.
in \cite{ShMMHarerZagier}. Description of matrix model correlators in terms of
the resolvents has two advantages: it provides the Ward identities
(\ref{virasoro_constraints}) in a simple form of the loop equations (\ref{loop_eq})
and it reveals the important hidden structure, the spectral curve. The drawbacks are
the divergency of series for the genus expansion and the lack of explicit formulas
for exact correlators.

The last two problems are resolved, e.g., by switching from the exact resolvents to
the Harer-Zagier functions, where the correlators are weighted with
additional factorial factors, i.e. by summing up the series expansion by the Pade method.

For $\beta \neq 1$ much less is known. So far we were able to obtain the
Harer-Zagier functions only for specific values of $\beta \neq 1$. Attempts to
evaluate, at least, the first $\beta - 1$ correction lead to some generalizations of
the hypergeometric equations which is a hint that something conceptual
needs to be done for the results to become simple for arbitrary $\beta$. Below our
preliminary results are summarized.

The one-point Harer-Zagier generating function is defined as
\eq {
    \phi (z) = \frac{4\beta}{\tau^2 - 1}\sum_{k=0}^\infty \sum_{N = 0}^\infty C_k \lb \frac{N}{\beta},\beta \rb \frac{z^k}{(2k-1)!!} \lb\frac{\tau-1}{\tau+1}\rb^N ,
}
where $C_k \lb \frac{N}{\beta},\beta \rb$ is the one-point correlator in
the $\beta$-deformed matrix model with matrix size equal to $N/\beta$.

\paragraph {The case of $\beta = 1$.}
\eq {
    \phi(\beta = 1, z, \tau) = \frac{1}{1-\tau z^2}
}
This is the classical result by J.Harer and D.Zagier.

It satisfies the differential equation derived from the integrability conditions
\eq {
    \lambda \frac{\p}{\p \lambda} \lb \frac{(1-\lambda)^2}{\lambda} \varphi(\lambda ,x)\rb = x \frac{\p}{\p x} \lb x^2 \varphi(\lambda ,x)\rb,
}
where $\varphi = \frac{\tau^2-1}{4} \phi$ and $\lambda = \frac{\tau-1}{\tau+1}$.

Now it turns out that the two- and three-point Harer-Zagier functions can be found
as well, and they are expressed through the arctangent function \cite{ShMMHarerZagier}, i.e.
remain elementary functions.

\paragraph {The case of $\beta = 2$.}
The Harer-Zagier function for $SO(N)$ matrix model has the form
\eq {
    \phi(\beta = 2, z, \tau) = \frac{\tau}{\tau - z -z^2 \tau} + \frac{\sqrt{z} (\tau+1)}{2} \frac{\mbox{arctan}\lb\frac{2\sqrt{z}(z\tau -z-1)\sqrt{\tau -z-z^2\tau}}{1 + (2-3\tau)z + (2-2\tau + 2\tau^2)z^2}\rb}{(\tau - z - z^2\tau)^{3/2}}
}
and it satisfies
\eq {
    \lsb \frac{2z+2\tau^2 z + \tau}{2 z} + (z-\tau + \tau^2 z)\frac{\p}{\p z}\rsb \phi(\beta = 2,z,\tau) = \frac{\tau}{2 z} + \frac{2 + 2\tau +2\tau^2 z + \tau^2}{2(2z+1)(1-z\tau)}
}
with the initial conditions $\phi(\beta = 2,z,\tau) = 1 + (\tau - 1) z + \dots$

\paragraph{The case of $\beta = 1/2$.}
The Harer-Zagier function for the Sp(N) matrix model has the form
\eq {
    \phi(\beta = 1/2, z,\tau) = \frac{1}{1-\tau z} + \sqrt{\frac{z}{1+\tau}}\ \frac{\mbox{arctan}\lb \frac{2\sqrt{z+\tau z}\sqrt{1-z\tau}}{2-z-2\tau z}\rb}{(1-\tau z)^{3/2}}
}
and it satisfies
\eq {
    \lsb \lb \frac{1}{z} - 3\tau - \frac{5}{2}\rb -z(1+\tau) \frac{\p}{\p z} + \frac{(1+\tau)(2-\tau z)}{z} \frac{\p}{\p \tau}\rsb \phi(\beta = 1/2, z,\tau) = \frac{1}{z}
}
with the initial conditions $\phi(\beta = 1/2, z,\tau) = 1 + (\tau + 1/2) z + \dots$

One can see that in both cases of $\beta = 2,1/2$, which correspond to classical
groups, the Harer-Zagier functions remain expressed in terms of arctangents. This,
however, is not the case in the general situation.

\paragraph {The case of $\beta = 3$.}
The Harer-Zagier function for $\beta = 3$ satisfies the differential equation

\eqna {
    & (1 + 8 z^2 \tau + 24 z^2 \tau^2 + 9 z^3 \tau - z \tau - 6 z - 33 z^2) \phi + (11 z^2 \tau  - 18 z^3 - 2 z + 9 z^4 \tau) \frac{\p \phi}{\p z} & \\ & + (9z-12z^2\tau +12 z^2 \tau^3  - 9 z \tau^2) \frac{\p \phi}{\p \tau} = 1 - 4 z - 4 z \tau \nonumber
}
at particular values of $z$ it becomes the hypergeometric equation and, hence, has
no solutions expressed in elementary functions. So, presumably, what one is searching
for is some clever deformation of the arctangent function from
the previously described cases.

We observe that as we move further and further away from $\beta = 1$, the complexity
of results increases. Further work is needed to clarify the situation.

\subsection {Identities for free energy}
It turns out that for $\beta \neq 1$ the Gaussian free energy has more structure
than one could expect.
\label{sec_free_energy}
\subsubsection{Definitions}
Let us define the partition function without
$\frac{1}{N!}$ factor. To avoid an ambiguity,
let us denote all the quantities in this normalization with tildes.

The partition function for the Gaussian model we are considering is
\eq {
    \wZ (N,\beta)= \int d \lambda_1 \dots d\lambda_N \prod_{i<j} \lb \lambda_i-\lambda_j\rb^{2\beta}
    e^{-\frac{1}{2g}\sum_i \lambda_i^2}=N!Z(N,\beta)
}
Instead of (\ref{48}) we now have
\eq {
    \wZ (N,\beta)= \sqrt{2\pi}^N \sqrt{g}^{\beta N^2 + (1-\beta)N} \prod_{k=1}^{N} \frac{\Gamma(1+\beta k)}{\Gamma(1+\beta)}
}
Defining the free energy as
\eq {
    \wF (N,\beta)= \ln \wZ \sim \sum_{k=1}^N \ln \Gamma \lb 1+\beta k\rb,
}
where the equivalence means equality up to terms quadratic and linear in the
matrix size $N$ (they can be absorbed into redefinition of $\beta$ and $g$).

\subsubsection{Difference equation}
Thus defined free energy satisfies a certain difference equation.
Consider
\eq {
    \wG (N,\beta) = \wF (N,\beta) - \wF (N -1,\beta) = \ln \Gamma \lb 1+\beta N \rb
}
then it is obvious that
\eq {
    \wG (N,\beta) - \wG \lb N - \frac{1}{\beta},\beta \rb = \ln \lb \beta N \rb
}
which implies that
\eq {
    \frac{\p}{\p N} \wG (N,\beta) - \frac{\p}{\p N} \wG \lb N - \frac{1}{\beta}, \beta \rb = \frac{1}{N}
    \label {free_energy_diff_eqn}
}

\subsubsection{Exact relation between $\mF_{SW}$ and $\wF$}
Comparison of (\ref{SW_diff_eqn}) and (\ref{free_energy_diff_eqn}) gives
\eq {
    \boxed{
        \mF_{SW} (N,\beta) = \wF \lb N - \frac{1}{\beta} ,\beta\rb = \sum_k^{N-\frac{1}{\beta}} \ln \Gamma(1+\beta k),
    }
}
so the only peculiarity is in change of upper limit of summation. In the
case of $\beta = 1$, it becomes $N - 1$ and acquires a clear physical meaning:
division of partition function by $N!$ implies that the eigenvalues are
indistinguishable bosons.

\subsubsection{Direct comparison of series}
However, it is still instructive not to appeal to this difference equation argument,
but to act straightforwardly and look directly at the perturbative expansions at large
$N$ in order to see, what one can deduce from them. One way to obtain these expansions
is to use the Euler-Maclaurin formula.

\paragraph{Euler-Maclaurin formula.}
We need this formula in the following form:
\eqna{
\frac{\p}{\p N}\sum_{k}^{N-1} f(k) =
f(N) - \frac{1}{2}f'(N) + \frac{1}{12}f''(N) - \frac{1}{720}f''''(N) - \ldots
= \sum_{m=0}^\infty \frac{B_{m}}{m!}\ \p^mf(N),
\label{EulerMcLaurinFormula}
}
where $B_m$ are the Bernoulli numbers, $\sum \frac{B_m}{m!}t^m = \frac{t}{e^t - 1}$.
The low limit in the sum is inessential,
as long as it does not depend on $N$.
In the following examples it is chosen to be $k=0$.

The first examples are
$$
\begin{array}{crl}
f(k) = 1 & \frac{\p}{\p N} N = 1 & \\
f(k) = k & \frac{\p}{\p N}\frac{N(N-1)}{2} = N - \frac{1}{2} & \\
f(k) = k^2 & \ \ \frac{\p}{\p N} \frac{N(N-1)(2N-1)}{6} = N^2 - N + \frac{1}{6}
& = N^2 -\frac{2N}{2} + \frac{2}{12} \\
f(k) = k^3 & \frac{\p}{\p N} \frac{N^2(N-1)^2}{4} =
N^3 - \frac{3}{2}N^2 + \frac{1}{2}N & = N^3 - \frac{3N^2}{2} + \frac{6N}{12} + 0 \\
\ldots && \\
\end{array}
$$

\paragraph{Different series as they are.}
Here all equalities are considered up to terms linear and constant in $N$ or $\Lambda$.

Summing up contributions from different genera, one gets
\eqna {
    \frac{\p}{\p N} \mF_{SW} \lb \frac{\Lambda}{g} \rb = &\frac{1}{g} \beta \Lambda \ln \Lambda +
    \frac{\beta-1}{2}\ln\Lambda + g\frac{1-3\beta+\beta^2}{12 \beta\Lambda} +
    g^2\frac{\beta - 1}{24\beta \Lambda^2} + g^3\frac{-1+5\beta^2-\beta^4}{360 \beta^3 \Lambda^3}
    +& \\ & + g^4\frac{1-\beta^3}{240\beta^3 \Lambda^4} + g^5\frac{2-7\beta^2-7\beta^4+2\beta^6}{2520 \beta^5 \Lambda^5} + o\lb\frac{1}{\Lambda^5}\rb  \nonumber
}

Expanding $F$ at various points one gets
\eqna {
\frac{\p}{\p N} \wF \lb \frac{\Lambda}{g} - \frac{1}{\beta} \rb = & \frac{1}{g}\beta \Lambda
\ln \Lambda + \frac{\beta-1}{2}\ln\Lambda + g\frac{1-3\beta+\beta^2}{12 \beta\Lambda} +
g^2\frac{\beta - 1}{24\beta \Lambda^2} + g^3\frac{-1+5\beta^2-\beta^4}{360 \beta^3 \Lambda^3} +
&  \nonumber\\ &+g^4\frac{1-\beta^3}{240\beta^3 \Lambda^4} +
g^5\frac{2-7\beta^2-7\beta^4+2\beta^6}{2520 \beta^5 \Lambda^5} +
o\lb\frac{1}{\Lambda^5}\rb &  \nonumber\\
\frac{\p}{\p N} \wF \lb \frac{\Lambda}{g} \rb = & \frac{1}{g}\beta \Lambda \ln \Lambda +
\frac{1+\beta}{2} \ln \Lambda + g\frac{1 + 3\beta +\beta^2}{12 \beta \Lambda} -
g^2\frac{1+\beta}{24 \beta\Lambda^2} - g^3\frac{1-5\beta^2+\beta^4}{360\beta^3 \Lambda^3} +&
\\ &+  g^4\frac{1+\beta^3}{240\beta^3\Lambda^4} +
g^5\frac{2-7\beta^2-7\beta^4+2\beta^6}{2520\beta^5\Lambda^5} + o\lb\frac{1}{\Lambda^5}\rb &
\nonumber\\
\frac{\p}{\p N} \wF \lb \frac{\Lambda}{g}-1 \rb = &\frac{1}{g}\beta\Lambda\ln\Lambda +
\frac{1-\beta}{2}\ln\Lambda +g\frac{1-3\beta +\beta^2}{12\beta\Lambda} +
g^2\frac{1-\beta}{24\beta\Lambda^2} - g^3\frac{1-5\beta^2+\beta^4}{360\beta^3\Lambda^3}+&
\nonumber\\ & + g^4\frac{\beta^3-1}{240\beta^3\Lambda^4} + g^5\frac{2 -7\beta^2-7\beta^4+2\beta^6}{2520\beta^5\Lambda^5} + o\lb\frac{1}{\Lambda^5}\rb & \nonumber
}

\paragraph {Interpretation.}

 From the above series, one can easily guess the following relations
\eq {
    \boxed{
    \mF_{SW} \lb \frac{\Lambda}{g},\beta \rb = \wF \lb \frac{\Lambda}{g} - \frac{1}{\beta},\beta \rb = - \wF \lb -\frac{\Lambda}{g}-1,\beta\rb = -\wF \lb \frac{\Lambda}{g},-\beta \rb,
    }
    \label{SW_Gauss_conj}
}

The first equality is expected to hold from our previous difference equation analysis.

It turns out that $\wF$ with shifted arguments also satisfies the same difference
equation. Indeed,
\eqna {
    -\wF \lb N ,-\beta\rb + \wF \lb N-1,-\beta \rb = -\ln\Gamma(1-\beta N)\ , \nonumber\\
    -\ln\Gamma(-\beta N) + \ln\Gamma(1-\beta N) = \ln (-\beta N)
}

For $\wF(-N-1,\beta)$ one has to assume that the lower limit of summation is less
than $-N-1$ (which is rather weird)
\eq {
    -\wF \lb -N - 1,\beta\rb + \wF \lb -N ,\beta \rb = -\ln\Gamma(1-\beta N)
}

Normally, shifting the expansion point for some function does not lead to series
similar to the initial one, but produces something which looks completely different.
The fact that this is not the case may be an indication that some not yet discovered
mathematical structure is present here. Perhaps, it is a peculiar feature of the
Gaussian potential or, may be such equalities have more general character. It is
interesting to see which of these unexpected identities survive generalization to non-Gaussian
eigenvalue models.


\begin{thebibliography}{12}

\bibitem{SWfirst}
 N.Seiberg and E.Witten,
Nucl.Phys., {\bf B426} (1994) 19-52, hep-th/9408099;
Nucl.Phys., {\bf B431} (1994) 484-550, hep-th/9407087

\bibitem{SW2} A.Klemm, W.Lerche, S.Theisen and S.Yankielowicz, Phys.Lett. {\bf B344}
(1995) 169-175,  arXiv:hep-th/9411048;\\
P.Argyres and A.Shapere, Nucl.Phys. {\bf B461} (1996) 437-459, hep-th/9509175

\bibitem{SW} A.Gorsky, I.Krichever, A.Marshakov, A.Mironov, A.Morozov,
Phys.Lett., {\bf B355} (1995) 466-477, hep-th/9505035;\\
R.Donagi and E.Witten, Nucl.Phys., {\bf B460} (1996) 299-334,
hep-th/9510101

\bibitem{SWrev} I.Krichever, hep-th/9205110; {\it Comm.Math.Phys.} {\bf 143}
(1992) 415;\\
B.Dubrovin, {\it Nucl.Phys.} {\bf B379} (1992) 627;\\
H.Itoyama and A.Morozov,
{\it Nucl.Phys.} {\bf B491} (1997) 529, hep-th/9512161;\\
A.Marshakov, {\sl Seiberg-Witten Theory and Integrable Systems}, World
Scientific 1999;\\
A.Gorsky and A.Mironov, hep-th/0011197

\bibitem{WDVV}E.Witten, Surv.Diff.Geom., {\bf 1} (1991) 243;\\
R.Dijkgraaf, E.Verlinde and H.Verlinde, Nucl.Phys.,
{\bf B352} (1991) 59;\\
A.Marshakov, A.Mironov and A.Morozov,
Phys. Lett. {\bf B389} (1996) 43, arXiv:hep-th/9607109;
Mod.Phys.Lett. {\bf A12} (1997) 773-787,  arXiv:hep-th/9701014;
Int.J.Mod.Phys. {\bf A15} (2000) 1157-1206,  arXiv:hep-th/9701123;\\
A.Mironov and A.Morozov, Phys.Lett., {\bf B424} (1998) 48-52,
hep-th/9712177\\
A.Veselov, hep-th/9902142;\\
H.W. Braden, A.Marshakov, A.Mironov and A.Morozov, Phys.Lett.,
 {\bf B448} (1999) 195, hep-th/9812078

\bibitem{MMbz}
A.Mironov and A.Morozov, 
JHEP {\bf 04} (2010) 040, arXiv:0910.5670;
J.Phys. {\bf A43} (2010) 195401, arXiv:0911.2396;\\
A.Popolitov, arXiv:1001.1407;\\
Wei He, Yan-Gang Miao,
Phys.Rev. {\bf D82} (2010) 025020, arXiv:1006.1214;\\
K.Maruyoshi and M.Taki,
arXiv:1006.4505;\\
F.Fucito, J.F.Morales, R.Poghossian and D. Ricci Pacifici,
arXiv:1103.4495;\\
Y.Zenkevich,
arXiv:1103.4843

\bibitem{MMM} A.Marshakov, A.Mironov and A.Morozov,
arXiv:1011.4491

\bibitem{LMNS}
G.Moore, N.Nekrasov, S.Shatashvili, Nucl.Phys. {\bf B534} (1998) 549-611, hep-th/9711108;
hep-th/9801061;\\
A.Losev, N.Nekrasov and S.Shatashvili, Commun.Math.Phys. {\bf 209} (2000) 97-121, hep-th/9712241;
ibid. 77-95, hep-th/9803265

\bibitem{NSh} N.Nekrasov and S.Shatashvili, arXiv:0908.4052

\bibitem{MMSh1} A.Mironov, A.Morozov, Sh.Shakirov,
JHEP {\bf 02} (2010) 030, arXiv:0911.5721

\bibitem{MMShtowaproof} A.Mironov, A.Morozov and Sh.Shakirov,
arXiv:1011.5629

\bibitem{AGT} L.Alday, D.Gaiotto and Y.Tachikawa,
Lett.Math.Phys. {\bf 91} (2010) 167-197, arXiv:0906.3219;\\
N.Wyllard,
JHEP {\bf 0911} (2009) 002, arXiv:0907.2189;\\
A.Mironov and A.Morozov,
Phys.Lett. {\bf B680} (2009) 188-194, arXiv:0908.2190;
Nucl.Phys. {\bf B825} (2009) 1-37, arXiv:0908.2569

\bibitem{AGTproof} L.Hadasz, Z.Jaskolski and P.Suchanek,
arXiv:0911.2353;
arXiv:1004.1841;\\
V.Fateev and I.Litvinov, JHEP {\bf 1002} (2010) 014, arXiv:0912.0504;\\
V.Alba, V.Fateev, A.Litvinov and G.Tarnopolsky,
arXiv:1012.1312;\\
A.Belavin and V.Belavin,
 arXiv:1102.0343

\bibitem{AGTmamo}
R.Dijkgraaf and C.Vafa, arXiv:0909.2453;\\
H.Itoyama, K.Maruyoshi and T.Oota,
Prog.Theor.Phys. {\bf 123} (2010) 957-987, arXiv:0911.4244;\\
T.Eguchi and K.Maruyoshi,
arXiv:0911.4797;
arXiv:1006.0828;\\
 R.Schiappa and N.Wyllard,
arXiv:0911.5337

\bibitem{AGTmamoproof}
H.Itoyama and T.Oota,
arXiv:1003.2929;\\
A.Mironov, A.Morozov and And.Morozov,
arXiv:1003.5752;\\
A.Mironov, A.Morozov and Sh.Shakirov,
JHEP {\bf 02} (2011) 067, arXiv:1012.3137

\bibitem{DV} R.Dijkgraaf and C.Vafa,
  Nucl.Phys. {\bf B644} (2002) 3, hep-th/0206255;
  Nucl.Phys. {\bf B644} (2002) 21, hep-th/0207106;
hep-th/0208048

\bibitem{DVmore}
L.Chekhov and A.Mironov,
  Phys.Lett. {\bf B552} (2003) 293, hep-th/0209085;\\
  A.Klemm, M.Marino and S.Theisen,
JHEP 0303 (2003) 051, hep-th/0211216;\\
   V.Kazakov and A.Marshakov,
  J.Phys.  {\bf A36} (2003) 3107, hep-th/0211236;\\
  H.Itoyama and A.Morozov,
Nucl.Phys.B657:53-78,2003, hep-th/0211245;
Phys.Lett. B555 (2003) 287-295, hep-th/0211259;
Prog.Theor.Phys. 109 (2003) 433-463, hep-th/0212032;
Int.J.Mod.Phys. A18 (2003) 5889-5906, hep-th/0301136;\\
L.Chekhov, A.Marshakov, A.Mironov and D.Vasiliev,
  Phys.Lett. {\bf B562} (2003) 323, hep-th/0301071;
Proc. Steklov Inst.Math. {\bf 251} (2005) 254,
hep-th/0506075;\\
A.Alexandrov, A.Mironov and A.Morozov,
Int.J.Mod.Phys. \textbf{A21} (2006) 2481-2518, hep-th/0412099;
Fortsch.Phys. \textbf{53} (2005) 512-521, hep-th/0412205;\\
A.Mironov, Theor.Math.Phys. \textbf{146} (2006) 63-72,
hep-th/0506158

\bibitem{AMMfirst}
A.Alexandrov, A.Mironov and A.Morozov,
Int.J.Mod.Phys. {\bf A19} (2004) 4127, hep-th/0310113

\bibitem{EOrev} B.Eynard and N.Orantin,
arXiv:math-ph/0702045

\bibitem{toprec}
A.Alexandrov, A.Mironov and A.Morozov,
Teor.Mat.Fiz. {\bf 150} (2007) 179-192, hep-th/0605171;
Physica {\bf D235} (2007) 126-167, hep-th/0608228; JHEP {\bf 12} (2009) 053,
arXiv:0906.3305;\\
A.Alexandrov, A.Mironov, A.Morozov, P.Putrov,
Int.J.Mod.Phys. {\bf A24} (2009) 4939-4998, arXiv:0811.2825;\\
B.Eynard,
JHEP \textbf{0411} (2004) 031, hep-th/0407261;\\
L.Chekhov and B.Eynard,
JHEP \textbf{0603} (2006) 014, hep-th/0504116;
JHEP \textbf{0612} (2006) 026, math-ph/0604014; \\
N.Orantin,
arXiv:0808.0635;\\
I.Kostov and N.Orantin, arXiv:1006.2028;\\
L.Chekhov, B.Eynard and O.Marchal,
arXiv:1009.6007

\bibitem{MMShDF} A.Mironov, A.Morozov, Sh.Shakirov,
Int.J.Mod.Phys. {\bf A25} (2010) 3173-3207, arXiv:1001.0563

\bibitem{GMAMO} A.Gerasimov, A.Marshakov, A.Mironov, A.Morozov and A.Orlov, Nucl.Phys.
{\bf B357} (1991) 565-618

\bibitem{Vir} A.A.~Migdal, 
Phys.Rep. {\bf 102} (1983) 199;\\
J.~Ambj{\o}rn, J.~Jurkiewicz and Yu.~Makeenko,
Phys.Lett. {\bf B251} (1990) 517;\\
F.~David, 
Mod.Phys.Lett. {\bf A5} (1990) 1019;\\
A.~Mironov and A.~Morozov, 
Phys.Lett. {\bf B252} (1990) 47-52\\
J.~Ambj{\o}rn and Yu.~Makeenko,
Mod.Phys.Lett. {\bf A5} (1990) 1753;\\
H.~Itoyama and Y.~Matsuo, 
Phys.Lett. {\bf 255B}
(1991) 202

\bibitem{confmm}
A.Marshakov, A.Mironov, and A.Morozov,
Phys.Lett. {\bf B265} (1991) 99\\
S.Kharchev, A.Marshakov, A.Mironov, A.Morozov and S.Pakuliak,
Nucl.Phys. {\bf B404} (1993) 17-750,  arXiv:hep-th/9208044;\\
H.Awata, Y.Matsuo, S.Odake and J.Shiraishi, Phys. Lett. {\bf B347} (1995) 49, hep-th/9411053;
Soryushiron Kenkyu {\bf 91} (1995) A69-A75, hep-th/9503028

\bibitem{ShMMCIV-DV}
A.Morozov and Sh.Shakirov,
arXiv:1004.2917

\bibitem{HZ} J.Harer and D.Zagier, Invent.Math. {\bf 85} (1986) 457-485;\\
S.K.Lando and A.K.Zvonkin, {\sl Embedded graphs}, Max-Plank-Institut f\"ur
Mathematik, Preprint Series 2001 ({\bf 63})

\bibitem{ShMMHarerZagier} A.Morozov and Sh.Shakirov,
JHEP {\bf 0912} (2009) 003 arXiv:0906.003;
arXiv:1007.4100

\end{thebibliography}
\end{document}